\documentclass[prb,preprint,floatfix,preprintnumbers]{revtex4}
\usepackage{graphicx}
\usepackage{dcolumn}
\usepackage{bm}
\begin{document}
\title{Prediction of magnetic moment collapse in ZrFe$_2$ under hydrostatic pressure}
\author{Wenxu Zhang, and Wanli Zhang}
\affiliation{State Key Laboratory of Electronic Thin Films and Integrated Devices, University of Electronic Science and Technology of China, Chengdu, 610054, P. R. China}
\date{\today}
\begin{abstract}
 Electronic structure and magnetic properties of ZrFe$_2$ with a cubic Laves phase are investigated by calculations based on the density functional theory. The total magnetic moment (m) of 3.14 $\mu_B$ per formula unit (\emph{f.u.}) is obtained at the experimental lattice constant (7.06 \AA), which is larger than 3.06 $\mu_B$/\emph{f.u.} obtained at the theoretical equilibrium lattice constant (6.85 \AA). The localized $3d$ magnetic moment is in negative diffusive sp background moment. We predict a two-step magnetic collapse under pressure: one is from 3.06 $\mu_B$/\emph{f.u.} to 1.26 $\mu_B$/\emph{f.u.} at about 3.6 GPa, and the other is from 0.5 $\mu_B$/\emph{f.u.} to nonmagnetic state at about 15 GPa.  We understand this process by the changes of density of states. The magnetic moment decreases under the pressure in the vicinity of the experimental lattice constant with $d\ln m/dp=-0.038$ GPa$^{-1}$. The spontaneous volume magnetostriction is 0.015. We suggest that the Invar effect of this alloy may be understood when considering the magnetic moment variation according to the Weiss $2\gamma$-model.
\end{abstract}
\maketitle
\section{Introduction}
Magnetic collapse, either being transition from ferromagnetic state to paramagnetic state, or from high spin (HS) state to low spin (LS) state, under pressure, is a widely observed phenomenon. Experiments, such as hyperfine field measurements\cite{pasternak},
X-ray magnetic dichroism\cite{duman}, nuclear forward scattering\cite{troyan}, can have a direct or indirect access to this phenomenon. Theoretical calculations based on density functional theory (DFT) were widely adopted to explain and predict it. For example, the HS-LS transition of transition metal monoxides (e.g. FeO, MnO, etc.) under the hydrostatic pressure as high as about 200 GPa were predicted by Cohen\cite{cohen}. Magnetic transition in these highly correlated insulators is the results of competition among the kinetic energy, exchange energy and Coulombic repulsion\cite{wxz}. The magnetic collapse in metals on the other hand can be qualitatively understood with the help of the Stoner model: In a simplified version of this model, a magnetic state is stable if $IN(E_F)>1$, where $I$ is the Stoner parameter, which is weakly dependent on the atomic distance, while $N(E_F)$, the density of states at the Fermi level, decreases as the band width increases under the pressure. At a certain critical pressure, the criterion is no longer satisfied, then the ferromagnetism cannot be sustainable. The magnetic moment collapse is highly expectable in ZrFe$_2$ under pressure when we compare the three isostructure compounds: YCo$_2$, ZrFe$_2$, and YFe$_2$, which have 93, 92, and 91 electrons, respectively. YCo$_2$ is metamagnetic, while YFe$_2$ is ferromagnetic with magnetic moments about 2.90 $\mu_B/f.u.$. Experiments towards this direction are not carried out according to our best knowledge. Here we approach it by calculations based on density functional theory.
\par Magnetism was proposed to be entangled with the Invar effect, where the material shows almost zero temperature dependence of the volume in a certain temperature region. The Invar effect is related to  the magnetic collapse by the fact that the thermal expansion of the lattice can be (partly) compensated by the decrease of the lattice constant induced by the decrease of the magnetic moment at the same time\cite{wasserman}.
\par As an Invar alloy, together with other interesting properties, ZrFe$_2$ and its doped versions were investigated by many researchers. Shiga\cite{shiga} reported the experimental evidence of the Invar effect in Laves phase intermetallic compounds, giving the spontaneous volume magnetostriction $\omega_s$ in ZrFe$_2$ being 0.01. The $\omega_s$ is defined in terms of the ratio of the equilibrium volumes in the ferromagnetic FM (V$_{FM}$) and the paramagnetic PM state (V$_{PM}$)
\begin{equation}
\omega_s=\frac{V_{FM}-V_{PM}}{V_{PM}}.
\end{equation}  
The pressure dependence of the Curie temperature(T$_c$) was measured by Brouha\cite{brouha}. T$_c$ was reported to be around
 625 K at the ambient pressure. A negative $dT_c/dp$ up to hydrostatic pressure of 35 kbar showed the characteristics of the strong ferromagnetism, as they proposed, in which only one spin subband is fully occupied. The pressure dependence of the hyperfine field (H$_{hf}$) at the Fe site of ZrFe$_2$ was measured up to 0.8 GPa by Dumelow\cite{dumelow}. The value of $d\ln H_{hf}/dp$ was $-7.3\pm0.1\times 10^{-4}$/kbar. The total magnetic moment in ZrFe$_2$ were measured by two authors as summarized in  Table~\ref{tab:table1}.
\par Klein \cite{klein} \emph{et al.} discussed the electronic structure, superconductivity, and magnetism in ZrX$_2$ (X=V, Fe, Co) with the C15 structures . Their results show that the Stoner theory is quantitatively inaccurate in these compounds because there is a significant covalent bonding. This bonding  mechanism in ZrFe$_2$ was proposed by Mohn\cite{mohn}. The consequence of this bonding is that the weights of DOSs of the majority and minority electrons also changes, rather than only a rigid shift of the subbands. The tiny energy difference between the paramagnetic and antiferromagnetic state at small lattices in their calculation has already indicated the magnetic collapse in ZrFe$_2$ under pressure, but no detailed information about the magnetic transition was given there\cite{mohn}. Recently, the Laves-phase alloy (Zr,Nb)Fe$_2$ was re-examined by Mohn\cite{mohn10}, where the connection of Invar with magnetic moment frustration were disentangled.
\par In this work, we investigate the magnetic behavior of ZrFe$_2$ under hydrostatic pressures by DFT. We show that the magnetic moment evolution envolves three steps: a continuous transition from a high spin (HS) state to a low spin (LS) state, followed by a discontinuous decrease to a metastable state (MS), and finally arriving at a nonmagnetic (NM) state under successively increasing the
pressure up to 26.0 GPa.
\section{Calculation details}
ZrFe$_2$ crystalizes in the C15 (space group F$d\bar{3}m$) structure Laves phase with two formula units per face centered cubic unit cell. The full-potential local orbital minimum basis band structure code (FPLO)\cite{fplo} was used in our calculation. The local spin density approximation of Perdew-Wang 92\cite{perdew} was adapted here. The number of k-points in the full Brillouin zone is $30\times 30\times 30$, which can guarantee the convergence of the total energy to microHartree. Both scalar relativistic and full-relativistic treatment were conducted and the results are compared. The fixed spin moment (FSM) calculations were used to investigate the possible multiple local energy minima with respect to the magnetic moment.
\section{Results and Discussions}
\subsection{magnetic moments at the equilibrium lattice constant}
The calculated properties and their comparison with the experimental and other theoretical ones are listed in Table~\ref{tab:table1}.
\begin{table}
\caption{\label{tab:table1} The calculated parameters of the cubic Laves phase ZrFe$_2$ compared with the experimental ones. The values without references are obtained in this work.}
\begin{ruledtabular}
\begin{tabular}{lcc}
parameters   & experimental        & calculational\\
\hline
a$_0$ (nm)          & 0.706\cite{warren} & 0.685, 0.698\cite{mohn}, 0.707\cite{yamada}\\
m$_{tot}$ ($\mu_B$)&3.46\cite{warren}, 3.14\cite{kanematsu}&  3.14, 3.24\cite{mohn}, 3.21\cite{yamada}\\
$\frac{d\ln{m}}{dp}$ (GPa$^{-1}$)&-&-0.038\\
$\omega_s$&0.01\cite{shiga} &0.015\\
\end{tabular}
\end{ruledtabular}
\end{table}

Our theoretical lattice constant (0.685 nm) by the scalar relativistic calculation  is about 3\% smaller than the experimental ones (0.706 nm by Warren\cite{warren}, 0.707 nm by Yamada \cite{yamada}). However, it is within the systematic error of L(S)DA, which usually underestimates the lattice constant\cite{barth}. The total magnetic moment obtained at the theoretical lattice is 3.06 $\mu_B/f.u.$, which is quite deviated from the experimental ones, while the calculation using the experimental lattice constants gives acceptable total magnetic moments of 3.14$\mu_B/f.u.$ by the full-relativistic calculations. The orbital moment of Fe atom is 0.054 $\mu_B$, and Zr 0.008 $\mu_B$ which compensates parts of the spin moment in the scalar relativistic case where the total spin moment are 2.60 $\mu_B/f.u.$ and 3.07 $\mu_B/f.u.$ at these two lattice constant, respectively. At equilibrium the body modulus and its derivative with respect to the pressure at the HS state are $B_0=49.41$ GPa and $B_0^\prime=4.75$ respectively by fitting the \emph{E-V} curve to the Birch-Murnaghan 3rd order equation of state
(EOS)\cite{birch}. The bulk modulus given by different fitting schemes, such as cubic polynomial, gives deviations of $\pm$2 GPa from the present value.
The NM state gives $B_0=56.94$ GPa and $B_0^\prime=4.167$, respectively.
\par After projecting the magnetic moment on different orbitals, we can observe that the 3d electrons of the two irons contribute about 4.08 $\mu_B$ to the total magnetic moment. Other orbitals have negative contributions. The magnetic moment of Zr consists of -0.4 $\mu_B$ from the 4d electrons and -0.15 $\mu_B$ from the 4sp electrons. This population analysis is in agreement with the
previous results by Mohn\cite{mohn}, who shows that there were diffusive negative moment background, which were from the d electrons, as well as the sp electrons.
\subsection{the magnetic collapse under the pressure}

\begin{figure}
\includegraphics[angle=0,scale=0.25]{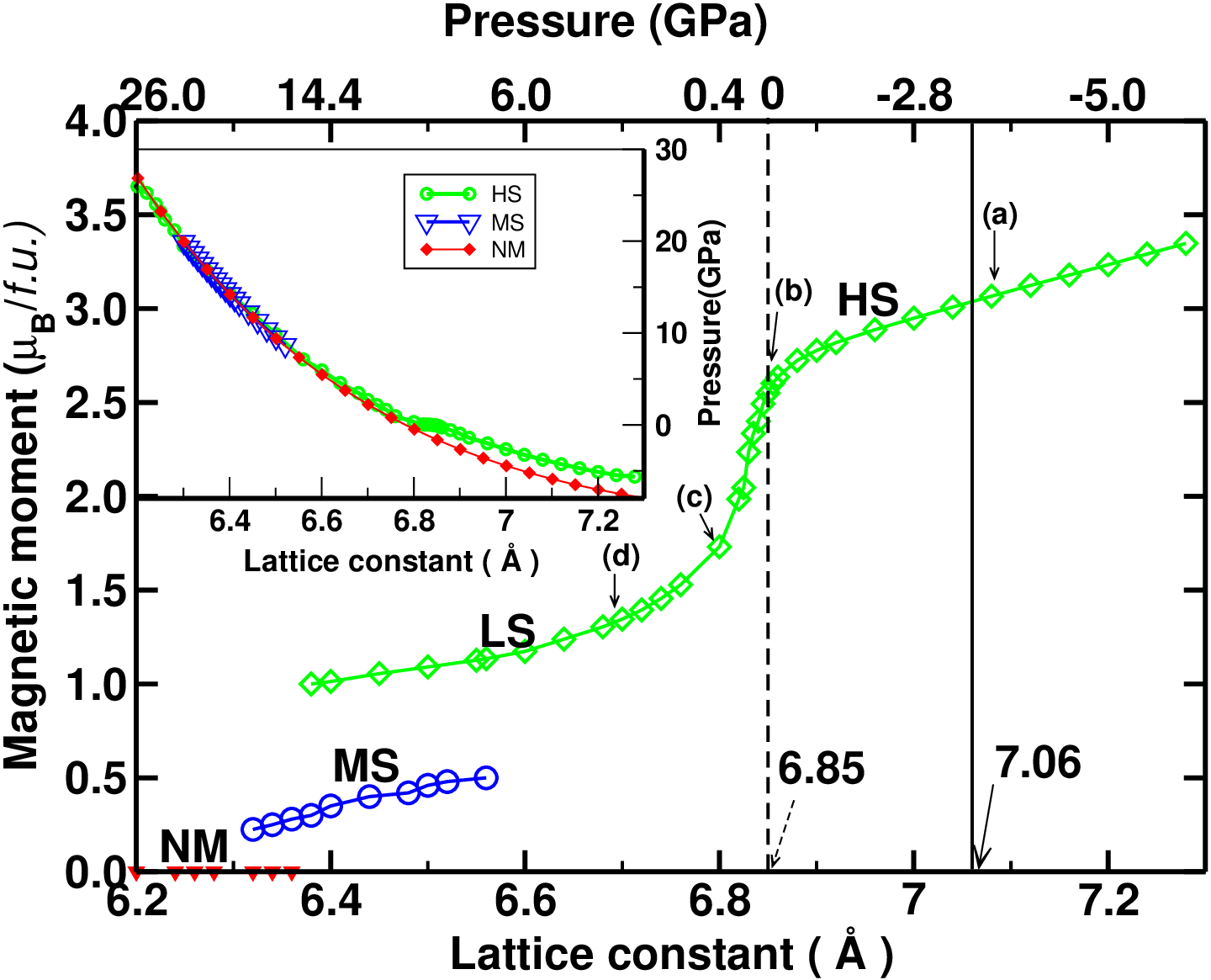}
\caption{\label{fig:mom-lat}(Color online) The magnetic moment evolution with the lattice constants. The experimental lattice constant is indicated by the solid vertical line, while the theoretical one by the dashed line. The inset gives the pressure-lattice relation of the three different states.}
\end{figure}
\begin{figure}
\vspace{0.5cm}
\includegraphics[scale=0.25]{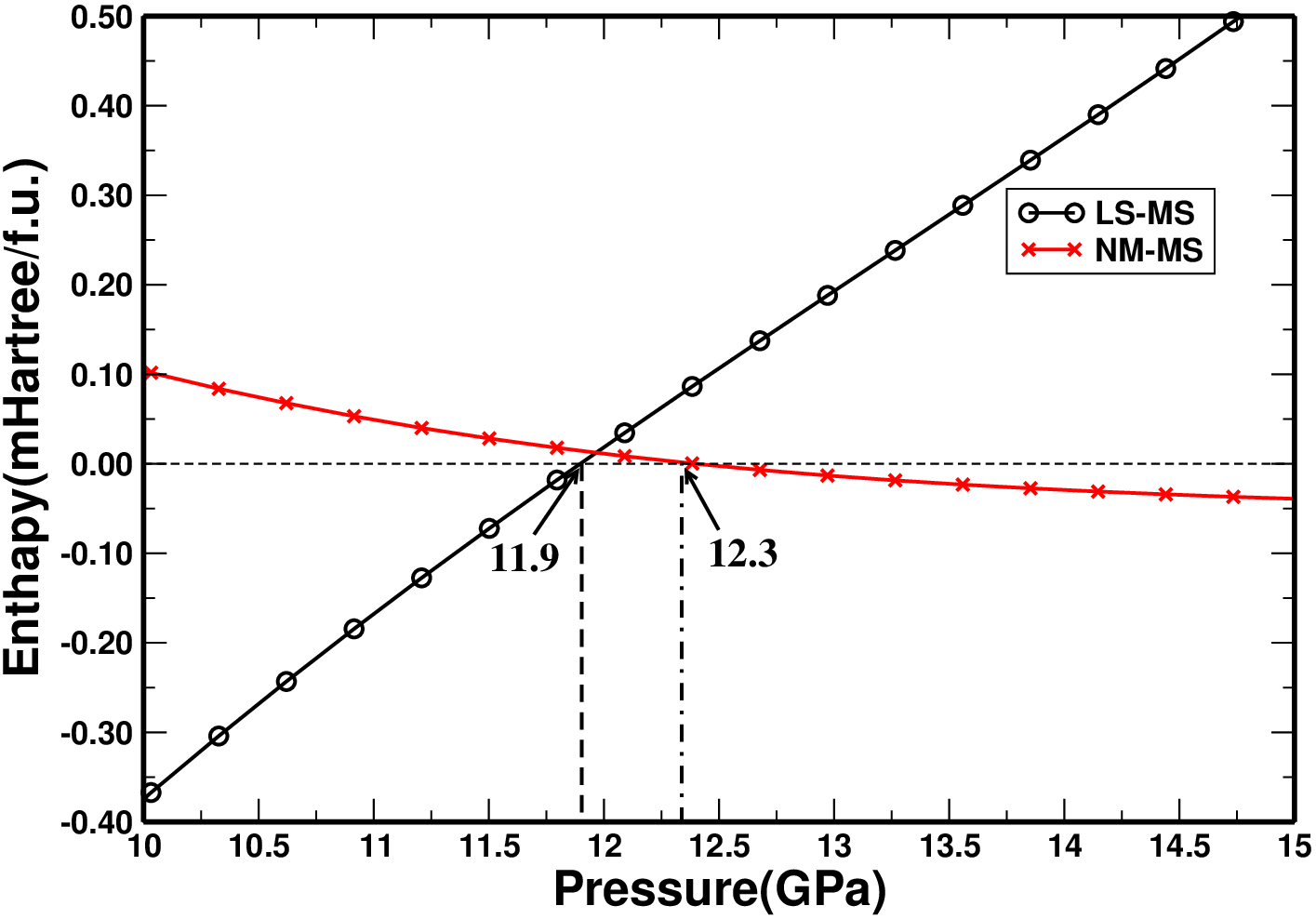}
\caption{\label{fig:enthapy}The enthalpy differences between the
\emph{LS} and \emph{MS} states, and \emph{MS} and \emph{NM} states under the pressure. The pressure is zero at the theoretical
equilibrium lattice constant in this figure. The pressure at the experimental equilibrium lattice is about -3 GPa.}
\end{figure}
\par The variation of total magnetic moment with the lattice constants is shown in Fig.~\ref{fig:mom-lat}. The corresponding hydrostatic pressures of different magnetic states are shown in the inset and also in the upper abscissa. Very obviously,
the magnetic moment decreases continuously from the larger value of 3.14 $\mu_B$ at $a_0=0.706$ nm to the smaller of 1.5 $\mu_B$ at
$a_0=0.676$ nm, corresponding to a hydrostatic pressure about 3.6 GPa. The pressure is calculated by taken the difference of the
pressure at experimental lattice constant and the one considered. Further compression of the lattice to about 0.657 nm, a metastable
state (MS) is initiated with the magnetic moment about 0.5 $\mu_B$ under the pressure about 6.0 GPa. Continuous decreasing the lattice to about 0.637 nm results that the LS state disappears. The phase transition can be shown by taken the differences of the enthalpy between the corresponding states. As shown in Fig.~\ref{fig:enthapy}, LS to MS transition takes places at 11.9 GPa, and MS to NM at 12.3 GPa. Total magnetic moment collapses at $a_0$ around 0.630 nm. The softening of the lattice due to the magnetism can be observed in the P-V curve in the inset of Fig.~\ref{fig:mom-lat}. The rather small magnetic moment about 0.2$\mu_B$/f.u. at the metastable state can possibly be suppressed by quantum fluctuation. Thus we anticipate a quantum phase transition of first order under pressure. When the magnetic moment disappears, superconductivity may be observed. This can be quite interesting, because according to our estimation the required pressure is about 15 GPa, which is readily available by the experiments.
\begin{figure}
\vspace{0.5cm}
\includegraphics[scale=0.32]{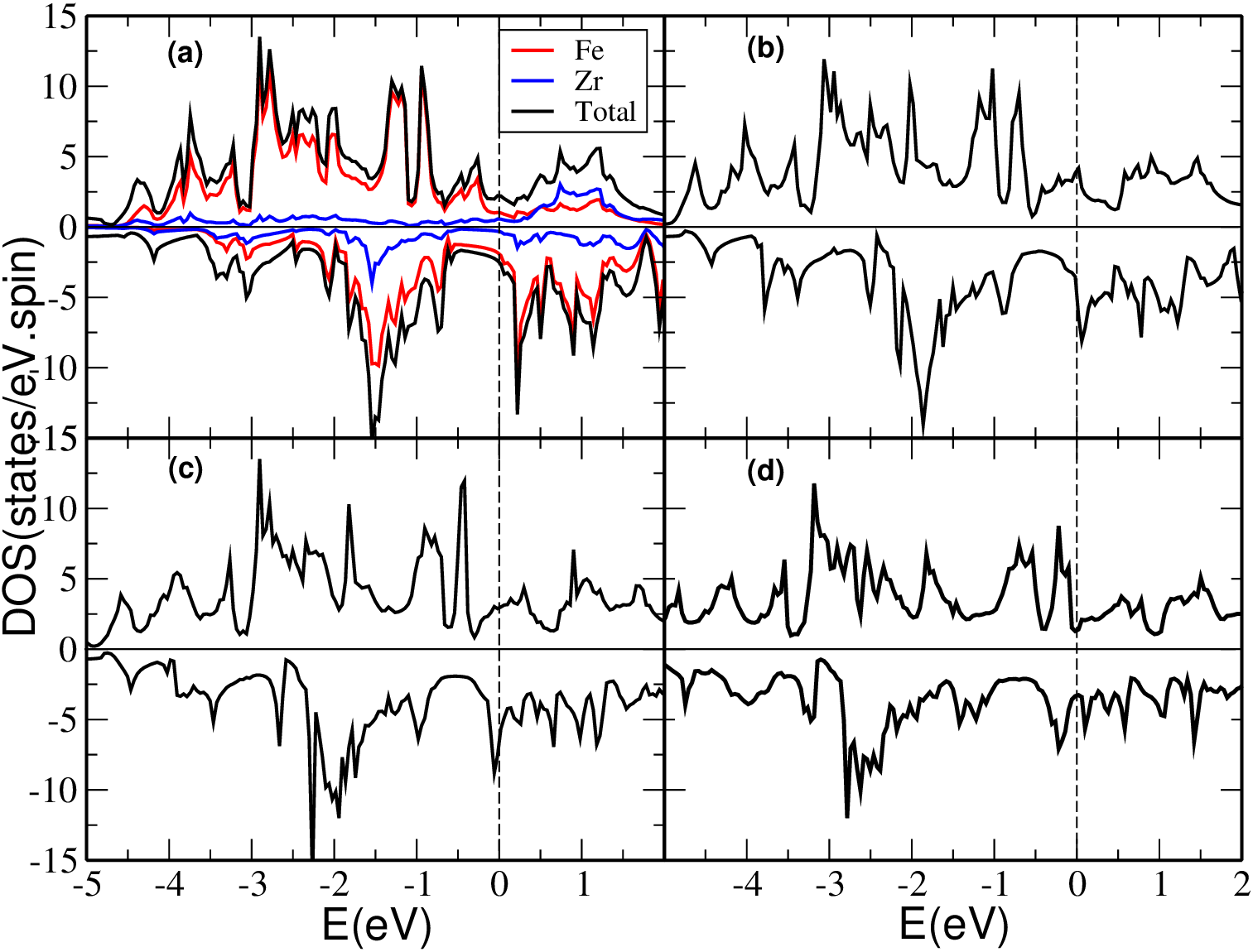}
\caption{\label{fig:dos}(Color online)The total DOSs and partial DOSs at different lattice constants. From (a) to (d), the lattice
constants are 7.08, 6.85, 6.80, and 6.70 \AA, respectively. The sign of the coordinate indicates the DOSs of the majority(+) and minority(-) spins. In (a), the partial DOSs of Fe and Zr are also shown.}
\end{figure}
\par The electronic origin of the magnetic moment behavior under the pressure can be explored by the DOSs. According to Fig.~\ref{fig:dos} (a$\sim$d), which show the DOS evolution with the lattice constants, the relative shift of the DOSs of the up and down spins is the reason for the decrease of the magnetic moment: the DOSs of the up spin electrons shift to higher energy while those of the down spin electrons shift to the lower energy when the lattice constant is reduced from (a) to (d).
\par The shape of the DOSs near the Fermi level (E$_F$) determines whether the magnetism collapses gradually or sharply. From 
Fig.~\ref{fig:dos}(a), it is observable that at the experimental lattice constant the DOSs of the up spin, contributed mainly from
the iron, has a gradual increase below the Fermi level, while the DOS of the down spin has a wide band dip about 0.8 eV below it and
a sharp increase just above it. The majority states of the Fe and Zr resonate above E$_F$, but the minority resonates below
E$_F$. This is the consequence of the covalent bonding. Thus an antiferromagnetic coupling between Fe and Zr is induced.
Applying the pressure will broaden the bandwidth due to the increase of the overlapping between the orbitals. This decreases the majority spin population while increases the minority spin electrons in order to conserve the total electron number. This process of gradual decrease of the magnetic moment is shown in Fig.~\ref{fig:mom-lat}, when the lattice constant is larger than 6.85 \AA. The decrease of the magnetic moment results in the decrease of the exchange field which is proportional to the magnetic moment. Thus the exchange splitting of the majority and minority spin is reduced. This shifts the high peak of the down spin nonbonding bands downward. When the Fermi level passes through the high DOS peak of the minority spins (Fig. 3(c)), the magnetic moment is rapidly reduced, as shown in Fig.~\ref{fig:mom-lat}, when the lattice constant is between 6.85 and 6.72 \AA. The occurrence of the MS state when a is between 6.32 and 6.58 \AA is due to the details of the DOSs. It is impossible to give an argument without calculations, but one thing is essential: A narrow peak around the Fermi level, so that the multiple magnetic solutions can exist\cite{wasserman}. In this lattice region, the material becomes weak ferromagnetism, with rather small magnetic moment: 0.1~$\mu_B$/Fe.

\subsection{The Invar effect in the compound}
More than twenty different models have been published in the past half century for understanding the Invar offect. A general review about the Invar can be found, for example, in the handbook edited by Buschow and Wohlfarth\cite{wasserman} and references therein.
One model called 2$\gamma$-model is based on the hypothesis of Weiss that there exist two separated energy minima with different
volumes and magnetic states: HS-high-volume and LS-low-volume states. First principle calculations by Entel\cite{entel} and other
authors supported this proposal. They argued that the special position of the Fermi level in the minority band, being at the
crossover between nonbonding and antibonding states, is responsible for the tendency of most Invar systems to undergo martensitic
phase transition. The HS-LS transition can also be continuous as proposed in this work in the cubic phase ZrFe$_2$, according to our calculation. The thermal excitation will cause the majority spins in the antibonding states flips to the minority nonbonding states. Increasing temperature, therefore, leads to a gradual loss of the spontaneous volume expansion associated with the ferromagnetic
state. This gradual process, contrary to the HS to LS in some Invar (e.g. Fe$_3$Pt), will not cause any discontinuity in the pressure
dependence of physical properties, which, to our knowledge, has not been observed experimentally. If we follow this, and estimate the
magnetovolume coupling constant $\kappa C$ by fitting the data around the transition region (a$_0$=6.7$\sim$6.85 \AA) to
$\omega_s=\kappa CM^2$, $\kappa C=1.7\times 10^{-8}cm^6emu^{-2}$ is obtained, which is comparable with the experimental value \cite{shiga} $\kappa C=2.2\times 10^{-8} cm^6emu^{-2}$. The overshooting of the spontaneous volume magnetostriction ($\omega_s$) and the underestimation of the magnetovolume coupling constant can partly because of the nonvanishing local magnetic moment above the
transition temperature.
\section{Conclusions}
\par In summary, we have shown that the pressure dependence of the magnetic moment in the cubic phase ZrFe$_2$ and elucidated it by the variation of the DOSs. The magnetic moment undergoes continuous transition from high spin state(3.14~$\mu_B/f.u.$) to low spin state (1.26~$\mu_B/f.u.$) at the hydrostatic pressure of 3.6 GPa, and further to lower spin state (0.5~$\mu_B/f.u.$) at about 14.9 GPa. The total magnetic moment collapses at the pressure about 15.3 GPa. We suggest that the Invar effect in this compound can be qualitatively understood by the spin flip transition due to the thermal excitation. We would also like to intrigue the experimentalist to investigate their volume (pressure) dependence of the magnetic and transport properties where the pressure are reachable in laboratories.
 \begin{acknowledgments}
 Discussions with M. Richter are greatly acknowledged. One of the authors, W.X. Zhang, would like to thank DAAD for the
financial support to study in Germany. Financial support from International Science \& Technology Cooperation Program of China (2012DFA51430) and Research Grant of Chinese Central Universities (ZYGX2013Z001) are acknowledged.
\end{acknowledgments}

\end{document}